\documentclass[cits]{PoS}

\vspace{-4.3cm}
\rightline{\parbox{12cm}{\hfill \rm\small DESY 07-170, Edinburgh 2007/32, Leipzig LU-ITP 2007/03, Liverpool LTH 763 }}           
\vspace{2.0cm}

\newcommand{\Dr}{\stackrel{\rightarrow}{D}}
\newcommand{\Drsl}{\not{\hspace{-0.1cm}{\Dr}}}

\newcommand{\Dslash}{\not{\hspace{-0.1cm}{D}}}
\newcommand{\pslash}{\not{\hspace{-0.08cm}p}}

\newcommand{\Aslash}{\not{\hspace{-0.08cm}A}}
\newcommand{\half}{{\textstyle \frac{1}{2}}}

\title{Perturbative determination of $c_{SW}$ with Symanzik improved gauge action and stout smearing}

\ShortTitle{Perturbative determination of csw with Symanzik improved gauge action and stout smearing}

\author{Roger Horsley\\
        School of Physics, University of Edinburgh, Edinburgh EH9 3JZ, UK\\
        E-mail: \email{rhorsley@ph.ed.ac.uk}}

\author{\speaker{Holger Perlt},
        ~Arwed Schiller\\
        Institut f\"ur Theortische Physik, Universit\"at Leipzig, PF 100 920, 04009 Leipzig, Germany\\
        E-mail: \email{Holger.Perlt@itp.uni-leipzig.de}\\
	E-mail: \email{Arwed.Schiller@itp.uni-leipzig.de}}

\author{Paul E.L. Rakow\\
        Theoretical Physics Division, Department of Mathematical Sciences,University of Liverpool,\\
	Liverpool L69 3BX, UK\\
        E-mail: \email{rakow@amtp.liv.ac.uk}}

\author{Gerrit Schierholz\\
        Deutsches Elektronen-Synchrotron DESY, 22603 Hamburg, Germany\\
        E-mail: \email{Gerrit.Schierholz@desy.de}}

\abstract{We determine the improvement factor $c_{SW}$ in one-loop lattice
perturbation theory for the plaquette and Symanzik improved gauge actions.
The fermionic action is ${\mathcal{O}(a)}$ clover improved with one-time stout smearing.
$c_{SW}$ is derived from the one-loop correction to the quark-quark-gluon vertex in the
off-shell regime. We give a first numerical value for the one-loop contribution to the non gauge-invariant
improvement coefficient $c_{NGI}$ for the quark field using the plaquette action. A discussion of mean field improvement
is included.
}

\FullConference{The XXV International Symposium on Lattice Field Theory\\
		 July 30 - August 4, 2007\\
		 Regensburg, Germany}

\begin{document}

\section{Introduction}

Current simulations with $2+1$ flavors require highly improved gauge and quark actions.
Renormalization group improved gauge actions are order  ${\mathcal{O}(a^2)}$ improved and 
should be preferred to the ${\mathcal{O}(a)}$ improved plaquette gauge action. In accordance
with our numerical simulations we take the Symanzik improved gauge action~\cite{Weisz:1982zw,Weisz:1983bn}
\begin{eqnarray}
S_G^{Symanzik} & = \frac{6}{g^2} \,\,\left[c_0
\sum_{plaq} \frac{1}{3}\, {\rm Re\, Tr\,}(1-U_{plaquette})
\, +  c_1 \sum_{rect} \frac{1}{3}\,{\rm Re \, Tr\,}(1- U_{
rectangle})\right]
\label{SG}
\end{eqnarray}
with
$$c_1=-\frac{1}{12}, \quad c_0=1-8\,c_1\,.$$
As the fermionic action, we use the clover improved action as proposed by
Sheikholeslami and Wohlert~\cite{Sheikholeslami:1985ij} which means that 
one has to add the so-called clover term to the standard Wilson fermion
action
\begin{eqnarray}
S_F^{clover}=S_F^{Wilson}- c_{SW}\,\sum_n\,\sum_{\mu,\nu}i g\, \frac{r}{4}\bar{\psi}_n\,\sigma_{\mu\nu}F_{\mu\nu}(n)\,\psi_n\,,
\label{SF}
\end{eqnarray}
where $F_{\mu\nu}(n)$ is the field strength in clover form and 
$\sigma_{\mu\nu}=i/2(\gamma_\mu\gamma_\nu-\gamma_\nu\gamma_\mu)$. An additional improvement can be achieved 
with ultraviolet filtering or smearing the gauge links $U_\mu$ in the fermionic Wilson action $S_F^{Wilson}$: 
it reduces the chiral
symmetry breaking of Wilson quarks among light flavors. There have been proposed several smearing techniques
(for a detailed discussion see~\cite{Capitani:2006ni}). We use the stout smearing of Morningstar
and Peardon~\cite{Morningstar:2003gk}. It is given by a sequence of transformations
\begin{equation}
	U_\mu \rightarrow U_\mu^{(1)} \rightarrow U_\mu^{(2)} \cdots \rightarrow U_\mu^{(n)}= \tilde{U}_\mu\, ,
\end{equation}
with
$$
	U_\mu^{(n+1)}(x) =  e^{iQ_\mu^{(n)} (U,\omega_{\mu\nu})}\,U_\mu^{(n)}(x)\,.
$$
The function $Q_\mu^{(n)} (U,\omega_{\mu\nu}) $ depends on the staples of the gauge link under consideration
and on the stout parameters $\omega_{\mu\nu}$ which determine the strength of smearing. We chose an isotropic
parameter $\omega_{\mu\nu}=\omega$ and one step smearing which is recommended by various investigations.

It is of importance to determine the improvement factor $c_{SW}$ appearing in (\ref{SF}) as precisely
as possible. Non-perturbative determinations are to be preferred but for the combination described
above there are no results obtained so far. In perturbation theory $c_{SW}$ has the form
\begin{equation}
c_{SW}=1 + g^2 \, c_{SW}^{(1)} + {\mathcal{O}(g^4)}\,.
\label{csw}
\end{equation}
There have been published results for $c_{SW}^{(1)}$ for plaquette action with twisted antiperiodic 
boundary conditions~\cite{Wohlert:1987rf} and Schr\"odinger functional method~\cite{Luscher:1996vw}.
For some popular improved gauge actions Aoki and Kuramashi~\cite{Aoki:2003sj} calculated the one-loop
correction using conventional perturbation theory. All results are obtained for unsmeared gauge links
in the on-shell regime.

In this paper we calculate $c_{SW}^{(1)}$ for Symanzik improved gauge action with stout smearing
in conventional perturbation theory. We do the calculation off-shell. This enables us 
to determine the one-loop contribution to the non gauge-invariant improvement coefficient $c_{NGI} $ for
the quark fields $\psi$ as proposed in~\cite{Martinelli:2001ak}. Using BRST symmetry arguments the authors proposed the off-shell
improvement for the quark fields $\psi_{\star}$ to be
\begin{equation}
\psi_{\star}=(1 + a \,c_D \Drsl + a \,i\,g\,\,c_{NGI} \Aslash)\psi\,,
\label{imppsi}
\end{equation}
where the coefficient $c_{NGI}$ does not contribute on-shell. 
Its perturbative expansion is known to be~\cite{Martinelli:2001ak}
\begin{equation}
c_{NGI}=g^2\,c_{NGI}^{(1)} + {\mathcal{O}(g^4)}\,.
\label{cNGI}
\end{equation}
In order to determine $c_{NGI}^{(1)}$ either a two-loop 
calculation of the quark propagator or a one-loop calculation of the quark-quark-gluon vertex is
required..
The improvement coefficient $c_D$ has been calculated to one-loop order in~\cite{Capitani:2000xi}.

\section{Improvement procedure}

In the approach of conventional perturbation theory we use the quark-quark-gluon vertex
$\Lambda_\mu(p_1,p_2)$ as discussed in~\cite{Aoki:2003sj} already. 
Looking at the ${\mathcal{O}}(a)$ expansion of tree-level $\Lambda^{(0)}_\mu(p_1,p_2)$ as derived from action (\ref{SF})
\begin{equation}
\Lambda^{(0)}_\mu(p_2,p_1) = -i\, g  \,\gamma_\mu -g\, \half \, a\, r\, {\bf 1} (p_1 + p_2)_\mu 
  - c_{SW} \,i\, g\, \half  a r \sigma_{\mu \alpha} (p_2 -p_1)_\alpha\\
  +\mathcal{O}(a^2)\,,
\label{treevertex}
\end{equation}
one can see by inserting (\ref{csw}) that a one-loop calculation for $\Lambda_\mu(p_2,p_1)$
provides necessary conditions to determine $c_{SW}^{(1)}$. 
We omit in all three-point functions the common overall color factor $t^{ac}$.
In (\ref{treevertex})
$p_1$ ($p_2$) are the incoming (outgoing) momenta. The off-shell improvement condition states that the
non-amputated improved three-point function $G_{\star,\mu}(p_2,p_1)$ has to be free of $\mathcal{O}(a)$ terms in one-loop.
The unimproved and improved three-point functions are defined by
\begin{eqnarray}
G_\mu(p_2,p_1)&=& S(p_2) \Lambda_\nu(p_2,p_1)S(p_1)D_{\nu\mu}(q)\, ,
\label{nonamp}
\\
G_{\star,\mu}(p_2,p_1)&=& S_\star(p_2) \Lambda_{\star,\nu}(p_2,p_1)S_\star(p_1)D_{\nu\mu}(q)\, ,
\label{nonampimp}
\end{eqnarray}
with $q=p_2-p_1$. $D_{\nu\mu}(q)$ is the full gluon propagator which is $\mathcal{O}(a)$-improved already. 
$\Lambda_{\mu}(p_2,p_1)$ and $\Lambda_{\star,\mu}(p_2,p_1)$ are the unimproved and
improved amputated three-point functions.
The corresponding quark propagators are given by
\begin{eqnarray}
S^{-1}(p)&=&i \pslash\, \Sigma_p(p) +\frac{ap^2}{2}\Sigma_W(p)=
	i \pslash \,\Sigma_p(p)\left(1-\frac{1}{2}a\, i \pslash\,\frac{\Sigma_W(p)}{\Sigma_p(p)}  \right)\,,
\label{S}
\\
S_\star^{-1}(p)&=&i \pslash\, \Sigma_p(p)\,.
\label{selfenergy}
\end{eqnarray}
In terms of the improved quark fields (\ref{imppsi}) $G_{\mu}(p_2,p_1)$  can
be related to its improved version
\begin{equation}
G_\mu(p_2,p_1)= G_{\star,\mu}(p_2,p_1) - 
	a \,i g\, c_{NGI}\, \mathcal{F}\left[\langle\left(\Aslash \Dslash^{-1} +  \Dslash^{-1}\Aslash  \right)A_\mu \rangle\right]\,.
\label{Gimp}
\end{equation}
In deriving (\ref{Gimp}) we have assumed $\langle A \rangle = 0$, $\mathcal{F}$ denotes the Fourier transform. Taking into account (\ref{cNGI}) we
insert in our one-loop calculation the corresponding tree-level expressions
\begin{equation}
a \,i g\, c_{NGI}\, \mathcal{F}\left[\langle\left(\Aslash \Dslash^{-1} +  \Dslash^{-1}\Aslash  \right)A_\mu \rangle^{tree}\right] =
a \,i g^3\, c_{NGI}^{(1)} \left(\gamma_\nu\frac{1}{i\pslash_1}+\frac{1}{i\pslash_2}\gamma_\nu\right)\,D^{tree}_{\nu\mu}(q)\,,
\label{cNGI1}
\end{equation}
or its amputated version
\begin{equation}
a \,i g\, c_{NGI}\, \mathcal{F}\left[\langle\left(\Aslash \Dslash^{-1} +  \Dslash^{-1}\Aslash  \right)A_\mu \rangle^{tree}_{amp}\right] =
-a \, g^3\, c_{NGI}^{(1)} \left(\pslash_2\gamma_\mu+\gamma_\mu\pslash_1\right)\,.
\label{cNGI1amp}
\end{equation}
If we amputate (\ref{nonamp}) and use (\ref{S}), (\ref{Gimp}) and  (\ref{cNGI1amp}) we get 
the off-shell improvement condition 
\begin{eqnarray}
\Lambda_{\mu}(p_2,p_1)&=&\Lambda_{\star,\mu}(p_2,p_1)+ 
	a \, g^3\, c_{NGI}^{(1)} (\pslash_2\gamma_\mu +\gamma_\mu\pslash_1)
\nonumber\\
& & -\frac{1}{2}a\,i\pslash_2 \frac{\Sigma_W(p_2)}{\Sigma_p(p_2)}\Lambda_{\star,\mu}(p_2,p_1)
     -\frac{1}{2}a\,i\,\Lambda_{\star,\mu}(p_2,p_1)\,\pslash_1 \frac{\Sigma_W(p_1)}{\Sigma_p(p_1)}\,,
\label{impcond}
\end{eqnarray}
which should hold to order $\mathcal{O}(g^3)$ by determining $c_{NGI}^{(1)}$ and $c_{SW}^{(1)}$
correctly.

\section{Calculation}

\begin{figure}[!htb]
  \begin{center}
      \includegraphics[scale=0.3,width=0.7\textwidth]{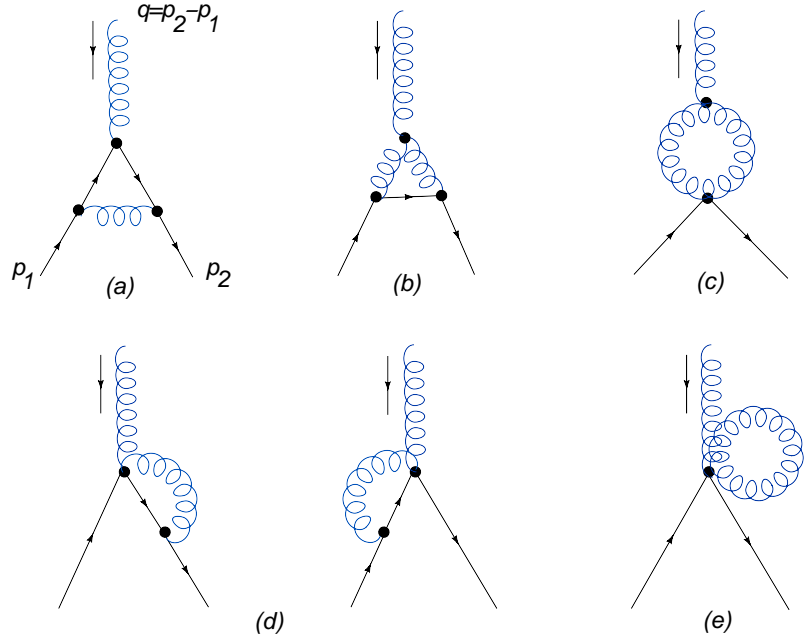}
        \end{center}
	\caption{One-loop diagrams contributing to the amputated quark-quark-gluon vertex}
\label{fig2}	
\end{figure}
The diagrams contributing to the amputated
one-loop three-point function are shown in Fig.~\ref{fig2}.
The calculation is performed combining symbolic and numerical methods. For the symbolic
computation we use a {\it Mathematica} package that we developed for one-loop calculations
in lattice perturbation theory (for a more detailed description see ~\cite{Gockeler:1996hg}). 
It is based on the infinite volume algorithm of 
Kawai et al.~\cite{Kawai:1980ja}. The analytic treatment has several advantages: one can extract
the infrared singularities exactly and the results are given as functions of lattice integrals 
which can be determined with high precision. The disadvantage consists in very large expressions
especially for the problem under consideration. In the analytic method the divergencies are 
isolated by differentiation with respect to external momenta. As can be seen in Fig. \ref{fig2}
diagrams (b) and (c) have two gluon propagators. So no parametrization can be chosen with only internal
momentum flowing through the gluon lines. Therefore at least one gluon propagator has to be differentiated.
Looking at the full analytic form of the gluon propagator for improved gauge actions~\cite{Horsley:2004mx}
one easily sees that huge analytic expression would arise. As discussed in~\cite{Horsley:2004mx}
one can split the full gluon propagator $D_{\mu\nu}^{{\rm improved}}(k)$ 
\begin{equation}
D_{\mu\nu}^{{\rm improved}}(k)=D_{\mu\nu}^{{\rm plaquette}}(k) + \Delta D_{\mu\nu}(k)\,.
\end{equation}
The diagrams with $D_{\mu\nu}^{{\rm plaquette}}(k)$ only contain the logarithmic parts and are treated
with the analytic {\it Mathematica} package. The diagrams with at least one $\Delta D_{\mu\nu}(k)$
are infrared finite and can be determined safely with pure numeric methods. We have written a
C program with a Gauss-Legendre integration algorithm in four dimensions 
(for a description of the method see~\cite{Gockeler:1996hg,Gockeler:2004xb}). We choose a sequence of small external
momenta $(p_1,p_2)$ and perform an extrapolation to vanishing momenta in order to extract the
corresponding values. Additionally, we have written an independent FORTRAN code which
computes the one-loop contributions for each diagram including the infrared logarithms. 
Results for both methods agree within accuracy.

The Feynman rules for non-smeared Symanzik gauge action have been summarized in~\cite{Aoki:2003sj}.
For the stout smeared gauge links in the clover action the rules are given for the forward
case by~\cite{Capitani:2006ni}. The corresponding Feynman rules needed for the quark-quark-gluon vertex are
much more complicated and have been derived by the authors. They are too long as to be given
in this proceedings~\cite{QCDSF2007}.

The calculation has been done in Feynman gauge with Wilson parameter $r=1$.
All the one-loop coefficients are calculated at $c_{SW}=1$ because
 $g^3 c_{SW} = g^3 + O(g^5)$.

\section{Results}

The anticipated general structure for the amputated three-point function in one-loop is
\begin{eqnarray}
\Lambda_\mu(p_2,p_1)&=& \Lambda^{{\overline{MS}}}_\mu(p_2,p_1)+A_{lat}\,i\,g^3\,\gamma_\mu
\nonumber\\
& & + B_{lat}\,\frac{a}{2}\,g^3\,\left(\pslash_2\,\gamma_\mu+\gamma_\mu\,\pslash_1\right)
+ C_{lat}\,\frac{i\,a}{2}\,g^3\,\sigma_{\mu\alpha}\,q_\alpha 
\end{eqnarray}
$\Lambda^{{\overline{MS}}}_\mu(p_2,p_1)$ is the universal part of the three-point function
independent of the chosen gauge action computed in the $\overline{MS}$-scheme
\begin{eqnarray}
\Lambda^{{\overline{MS}}}_\mu(p_2,p_1)&=& -i\, g\, \gamma_\mu - 
	g\, \frac{a}{2}\,{\bf 1}\left( p_{1,\mu}+p_{2,\mu}\right)-
        c_{SW}\,i\, g\,\frac{a}{2}\sigma_{\mu\alpha}\,q_\alpha
\nonumber\\
& & + i\, g^3\,F_{1,\mu}(p_1,p_2,q) + a\,g^3 \,F_{2,\mu}(p_1,p_2,q)\,.
\label{LamMS}
\end{eqnarray}
$F_{1,\mu}(p_1,p_2,q)$ and $F_{2,\mu}(p_1,p_2,q)$ are complicated functions involving polylogarithms
and logarithms. They will be given in~\cite{QCDSF2007}. The quantitites
$A_{lat}$, $B_{lat}$ and $C_{lat}$ are obtained as
\begin{eqnarray}
A_{lat}&=&C_F\,\left(0.03783 - 0.93653\,\omega + 3.42833\,\omega^2 + 0.01266\,\log (a\mu)   \right)
\nonumber\\
& &+
N_c\,\left(-0.02200  + 0.01266\,\log (a\mu)   \right)\,,
\nonumber\\
B_{lat}&=&C_F\,\left(0.03804 - 1.03749\,\omega + 3.43791\,\omega^2 + 0.02533\,\log (a\mu)   \right)
\nonumber\\
& &+
N_c\,\left(-0.02432 + 0.01925\,\omega  + 0.01266\,\log (a\mu)   \right)\,,
\\
C_{lat}&=&C_F\,\left(0.11618 + 0.82813\,\omega - 2.45508\,\omega^2\right)
\nonumber\\ 
& &+
N_c\,\left(0.01215 + 0.01109\,\omega - 0.30228\,\omega^2\right)\,,
\nonumber
\label{ABC}
\end{eqnarray}
with $C_F=(N_c^2-1)/(2N_c)$ for $SU(N_c)$. As shown in (\ref{impcond}) we need 
the self energy parts $\Sigma_p(p)$ and  $\Sigma_W(p)$ as defined in (\ref{S}) to solve the
off-shell improvement condition
\begin{eqnarray}
\Sigma_p(p)&=&1-\frac{g^2\,C_F}{16\pi^2}\left[\log (ap)^2 +\Sigma_1\right]\,,\nonumber\\
\Sigma_W(p)&=&1-\frac{g^2\,C_F}{16\pi^2}\left[2\,\log (ap)^2 +\Sigma_2\right]\,.
\label{SigmapW}
\end{eqnarray}
It turns out that the self energy parts $\Sigma_1$ and $\Sigma_2$ 
contribute only to $c_{NGI}^{(1)}$.
For the Symanzik gauge action we will present them in~\cite{QCDSF2007}.
For the plaquette action we get
\begin{eqnarray}
\Sigma^{plaq}_1 &=& 8.20627 -   196.44600\,\omega + 739.68364\,\omega^2\,,\nonumber\\
\Sigma^{plaq}_2 &=& 7.35794 -   208.58321\,\omega + 711.56526\,\omega^2\,.
\label{sigmas}
\end{eqnarray}
We use (\ref{LamMS}) and (\ref{ABC}) to construct the left hand side of (\ref{impcond})
whereas (\ref{SigmapW}) with (\ref{sigmas}) are inserted into the right hand side.
In order to fulfill (\ref{impcond}) we get the following improvement coefficients for the plaquette action
\begin{eqnarray}
c_{NGI}^{(1,plaq)}&=& 
N_c\,\left(0.00143 - 0.01166\,\omega \right)\,,
\\
c_{SW}^{(1,plaq)}&=&C_F\,\left(0.16764 + 1.07915\,\omega - 3.68668\,\omega^2\right)
\nonumber\\ 
& &+\,
N_c\,\left(0.01502 + 0.00962\,\omega - 0.28479\,\omega^2\right)\,.
\label{cpresults}
\end{eqnarray}
For the Symanzik improved gauge action we find the improvement coefficient $c_{SW}^{(1)}$
\begin{eqnarray}
c_{SW}^{(1)}&=&C_F\,\left(0.11618 + 0.82813\,\omega - 2.45508\,\omega^2\right)
\nonumber\\ 
& &+\,
N_c\,\left(0.01215 + 0.01109\,\omega - 0.30228\,\omega^2\right)\,.
\label{cresults}
\end{eqnarray}

\section{Mean field improvement}

It is known that lattice artefacts make the perturbative expansion worse. One possible improvement
procedure is to replace the naive coupling constant $g$ by its mean field improved value
$g_{MF}=g/u_0^2$ where $u_0^4$ is the average plaquette value for the corresponding gauge field action.
By scaling all gauge links in the clover field strength $F_{\mu\nu}(n)$ in (\ref{SF}) by $1/u_0$
one obtains the mean field improved $c_{SW}$ as
\begin{equation}
c_{SW}^{MF}=u_0^3\, c_{SW}\,.
\label{mf1}
\end{equation}
The perturbative expansion of $u_0$ is known to be
\begin{equation}
u_0=1-\frac{g_{MF}^2 C_F}{16\pi^2}\,k_u\,,
\label{mf2}
\end{equation}
where $k_u$ for popular gauge actions are given in~\cite{Horsley:2004mx}. Therefore, the perturbative
expression for the mean field improved $c_{SW}$ is given by
\begin{equation}
c_{SW}=c_{SW}^{MF}\,u_0^{-3}=\frac{1}{u_0^3}\,\left(1+g_{MF}^2\left(c_{SW}^{(1)}-\frac{3C_F}{16\pi^2}\,k_u\right)+
\mathcal{O}(g_{MF}^4)\right)=c_{SW}^{MF,p}+\mathcal{O}(g_{MF}^4)\,.
\label{mf3}
\end{equation}

For the future simulations of the QCDSF collaboration we have the following numbers 
for the Symanzik action and 2+1 flavors
$$
C_F=4/3,\quad  N_c=3, \quad u_0^4=0.6065, \quad g_{MF}^2=1.71335, \quad k_u=0.732524\,\pi^2\,.
$$
This gives the one-loop expression for $c_{SW}$ parameter as
\begin{eqnarray}
c_{SW}&=&1+g^2\,(0.19136 + 1.13745\,\omega -  4.18029\, \omega^2)+\mathcal{O}(g^4)\,,
\label{cswres}\\
c_{SW}^{MF,p}&=&\frac{1}{u_0^3}\left(1+g_{MF}^2\,(0.19136 + 1.13745\,\omega -  4.18029\, \omega^2)
	-g_{MF}^2\,0.18313\right)
\nonumber \\
&= & 1.47557 + 2.83568 \,\omega - 10.42148 \,\omega^2
\label{cswMFres}
\end{eqnarray}
For no stout-smearing ($\omega=0$) the result (\ref{cswres}) has to be compared with the number
given in~\cite{Aoki:2003sj}: $c_{SW}^{(1,AK)}=0.19624449(1)$. 
The minor difference to our value $c_{SW}^{(1)}=0.19136$ can possibly be related to an
 inaccuracy in our numerical integrations.
In the simluation the stout parameter $\omega$ is chosen to be $\omega=0.1$ leading to
a mean field improved value $c_{SW}^{MF,p}=1.65492$.

\vspace{1cm}

This investigation has been  supported by the DFG under contract FOR 465 (Forschergruppe 
Gitter-Hadronen-Ph\"anomenologie).

\end{document}